\documentclass[aps,10pt,prl,twocolumn,letterpaper,superscriptaddress]{revtex4-1}

\usepackage{graphicx} 
\usepackage{bm} 
\usepackage{amsmath}

\usepackage{xcolor}

\definecolor{PRLblue}{rgb}{0.18,0.18,0.57}
\usepackage[colorlinks=true,allcolors=PRLblue]{hyperref} 
\graphicspath{ {./}{figs/}}
\newcommand{\myhash}{\scalebox{0.8}{\raisebox{0.4ex}{\#}}}

\usepackage{fancyhdr}
\begin{document}
\title{Electronic structure and topology across $T_c$ in magnetic Weyl semimetal Co$_3$Sn$_2$S$_2$ }

\author{Antonio Rossi}
\thanks{These two authors contributed equally}
\affiliation{Department of Physics and Astronomy, University of California, Davis, CA 95616, USA}
\affiliation{Advanced Light Source, Lawrence Berkeley National Lab, Berkeley, 94720, USA}

\author{Vsevolod Ivanov}
\thanks{These two authors contributed equally}
\affiliation{Department of Physics and Astronomy, University of California, Davis, CA 95616, USA}

\author{Sudheer Sreedhar}
\affiliation{Department of Physics and Astronomy, University of California, Davis, CA 95616, USA}

\author{Adam L. Gross}
\affiliation{Department of Physics and Astronomy, University of California, Davis, CA 95616, USA}

\author{Zihao Shen}
\affiliation{Department of Physics and Astronomy, University of California, Davis, CA 95616, USA}

\author{Eli Rotenberg}
\affiliation{Advanced Light Source, Lawrence Berkeley National Lab, Berkeley, 94720, USA}

\author{Aaron Bostwick}
\affiliation{Advanced Light Source, Lawrence Berkeley National Lab, Berkeley, 94720, USA}

\author{Chris Jozwiak}
\affiliation{Advanced Light Source, Lawrence Berkeley National Lab, Berkeley, 94720, USA}

\author{Valentin Taufour}
\affiliation{Department of Physics and Astronomy, University of California, Davis, CA 95616, USA}

\author{Sergey Y. Savrasov}
\affiliation{Department of Physics and Astronomy, University of California, Davis, CA 95616, USA}

\author{Inna M. Vishik}
\email{ivishik@ucdavis.edu}
\affiliation{Department of Physics and Astronomy, University of California, Davis, CA 95616, USA}

\date{\today}

\begin{abstract}	
Co$_3$Sn$_2$S$_2$ is a magnetic Weyl semimetal, in which ferromagnetic ordering at 177K is predicted to stabilize Weyl points.  We perform temperature and spatial dependent angle--resolved photoemission spectroscopy measurements through the Curie temperature ($T_c$), which show large band shifts and renormalization concomitant with the onset of magnetism. 
We argue that Co$_3$Sn$_2$S$_2$ evolves from a Mott ferromagnet below $T_c$ to a correlated metallic state above $T_c$.
To understand the magnetism, we derive a tight--binding model of Co-$3d_{x^2-y^2}$ orbitals on the kagome lattice. At the filling obtained by first--principles calculations, this model reproduces the ferromagnetic ground state, and results in the reduction of Coulomb interactions due to cluster effects. 
Using a disordered local moment simulation, we show how this reduced Hubbard-$U$ leads to a collapse of the bands across the magnetic transition, resulting in a correlated state which carries associated characteristic photoemission signatures that are distinct from those of a simple lifting of exchange splitting.
The behavior of topology across $T_c$ is discussed in the context of this description of the magnetism.
\end{abstract}

\maketitle

\section{I. Introduction.}

Quantum materials encompass a broad array of phenomena, unified by the concept of emergent phases \cite{Keimer_quantum_materials_2017,Tokura_EmergentQuantumMaterials_2017}, which can arise as result of thermodynamic or topological phase transitions. Recently, magnetic topological materials have united these two paradigms with a promise to realize novel elementary excitations in a switchable manner, as exemplified by magnetic Weyl semimetals (WSMs). Magnetic WSM states were initially predicted in rare--earth pyrochlore iridates \cite{Savrasov_arcs}, and HgCr$_2$Se$_4$ \cite{HgCr2Se4-wsm}, and recently observed in a number of materials\cite{Belopolski_WeylDrumhead_2019, Cai_magWeyl_2020, Chang_MagneticWeylRAlGe_2018, Chang_MagneticWeylRAlGe_2018, Kuroda_MagneticWeyl_2017, Takiguchi_WeylFermion_SRO_2020}.

A WSM state occurs as an intermediate metallic regime between normal and topologically insulating phases when either inversion ($\mathcal{I}$) or time--reversal ($\mathcal{T}$) symmetries are broken \cite{WSMRMP, Wehling_DiracMaterials_2014, Yan2017}. Specifically, magnetic WSMs break $\mathcal{T}$ symmetry, and can realize the minimal number of Weyl points in the Brillouin zone (BZ), since the preserved inversion symmetry guarantees the existence of Weyl points with opposite chirality at $\bm{k}/\bm{-k}$ momenta, necessarily satisfying the Nielsen-Ninomiya theorem \cite{Nielsen}. This chirality is topological property that manifests as Fermi arc surface states \cite{Savrasov_arcs, WSMRMP} connecting pairs of projected Weyl points, and generally guarantees their stability, since the chirality can only be removed by merging with a Weyl point of opposite chirality.



A prototype magnetic WSM is Co$_3$Sn$_2$S$_2$, which is a ferromagnet with a Curie temperature of $T_c= 177$K \cite{Kubodera2006,Schnelle2013}. The structure is composed of triangular S and Sn layers interspersed between the Kagome lattice planes of Co atoms that are responsible for ferromagnetic order. This material possesses an unusually large anomalous Hall conductivity as a direct consequence of its topological band structure \cite{Liu2018,Li2020}. Other indications of the topology in Co$_3$Sn$_2$S$_2$ are the chiral anomaly-related negative longitudinal magnetoresistance \cite{Liu2018} and gigantic magneto-optical response \cite{Okamura2020}.  The topological nature of the band structure has been verified spectroscopically at low temperature \cite{Morali2019, liu_magnetic_2019,jiao_signatures_2019}, but limited spectroscopic information is available through $T_c$ \cite{yang_magnetization-induced_2020,Holder_PhotoemissionCo3Sn2S2_2009}. 

Switchable topology in magnetic WSMs has been implied through various experiments, including measurements of the intrinsic anomalous Hall \cite{HgCr2Se4-wsm, Shi_WSM_ferrimagnet_prediction_2018, Wang2018, Liu2018, Thakur2020} and Nernst effects \cite{Guin2019, Sakai2018, Asaba2021}, but direct signatures of band structure changes across $T_c$ have been less explored. This is in part because the microscopic details of magnetism can be quite complex \cite{NatComm-DMFT,yang_magnetization-induced_2020, Felser2021_domains,Kassem2017,Lachman2020,Guguchia2020}, resulting in a non--obvious evolution of the band structure and its topological features.
%
%
The topological physics of Co$_3$Sn$_2$S$_2$ is simultaneously manageable due to it only possessing six Weyl points, the minimum allowed by the combined $\mathcal{I}$ and $C_3$ symmetries, yet is still highly non-trivial because its magnetic structure is complicated by correlations \cite{NatComm-DMFT, yang_magnetization-induced_2020, Kassem2017}, frustrations \cite{Lachman2020,Guguchia2020}, and domains \cite{Felser2021_domains}.
%
%
These aspects make this material a convenient experimental platform for observing an elementary property of WSM: the merging and annihilation of Weyl points with opposite chirality across a magnetic phase transition \cite{zrte5-weylcancel, EuB6-prx}. 
Studying this theoretically predicted topological property of Weyl points holds not only pedagogical importance, but is also a crucial step towards controlling and manipulating electronic states in this system.



In the present work, we investigate the evolution of electronic structure across $T_c$ in magnetic WSM Co$_3$Sn$_2$S$_2$ using spatial and temperature-dependent angle-resolved photoemission spectroscopy (ARPES) and corresponding first principles calculations. Our ARPES measurements reveal a substantial band shift and renormalization across $T_c$, as well as a second band farther from the Fermi energy.
We draw an analogy from the well--known Mott--Hubbard physics, explaining this result as the emergence of a moderately correlated metallic state above $T_c$, characterized by a distinctive renormalized quasiparticle band near the Fermi energy, with the persistent spectral weight attributed to the remnants of the lower Hubbard band.
To better understand the evolution across $T_c$, we perform calculations based on the Disordered Local Moment (DLM) \cite{Pindor-DLM} picture of magnetism to show how persisting local magnetic correlations can lead to a renormalized quasiparticle band.
We then show that the states associated with the bulk Weyl nodes and Fermi arcs are absent above $T_c$, consistent with the reappearance of $\mathcal{T}$ symmetry. 
However, in our proposed model the persistent local moments above $T_c$ may also break the translational symmetry required for Weyl nodes to exist, oppugning the notion that they annihilate across the magnetic transition, and emphasizing the crucial role played by many-body effects during the evolution of magnetism and topology across $T_c$.

This paper is organized as follows: in Section II we discuss the magnetism of Co$_3$Sn$_2$S$_2$, and derive a minimal tight binding model, while Section III focuses on understanding the surface terminations of Co$_3$Sn$_2$S$_2$. The remainder of the work illustrates the changes in the band structure on the Sn termination across $T_c$, by considering the magnetic disorder and topological features in this material. In Section IV, the derived tight binding model is used in combination with first--principles disordered local moment simulations to understand the ARPES data across $T_c$. Section V then depicts the evolution of bulk Weyl cones across the magnetic transition, and uses the results of slab calculations to describe how the associated Fermi arc states will evolve. 

\section{II. Mechanism of Magnetic Order.}

\begin{figure}[ht]
	\includegraphics[width=1.0\columnwidth]{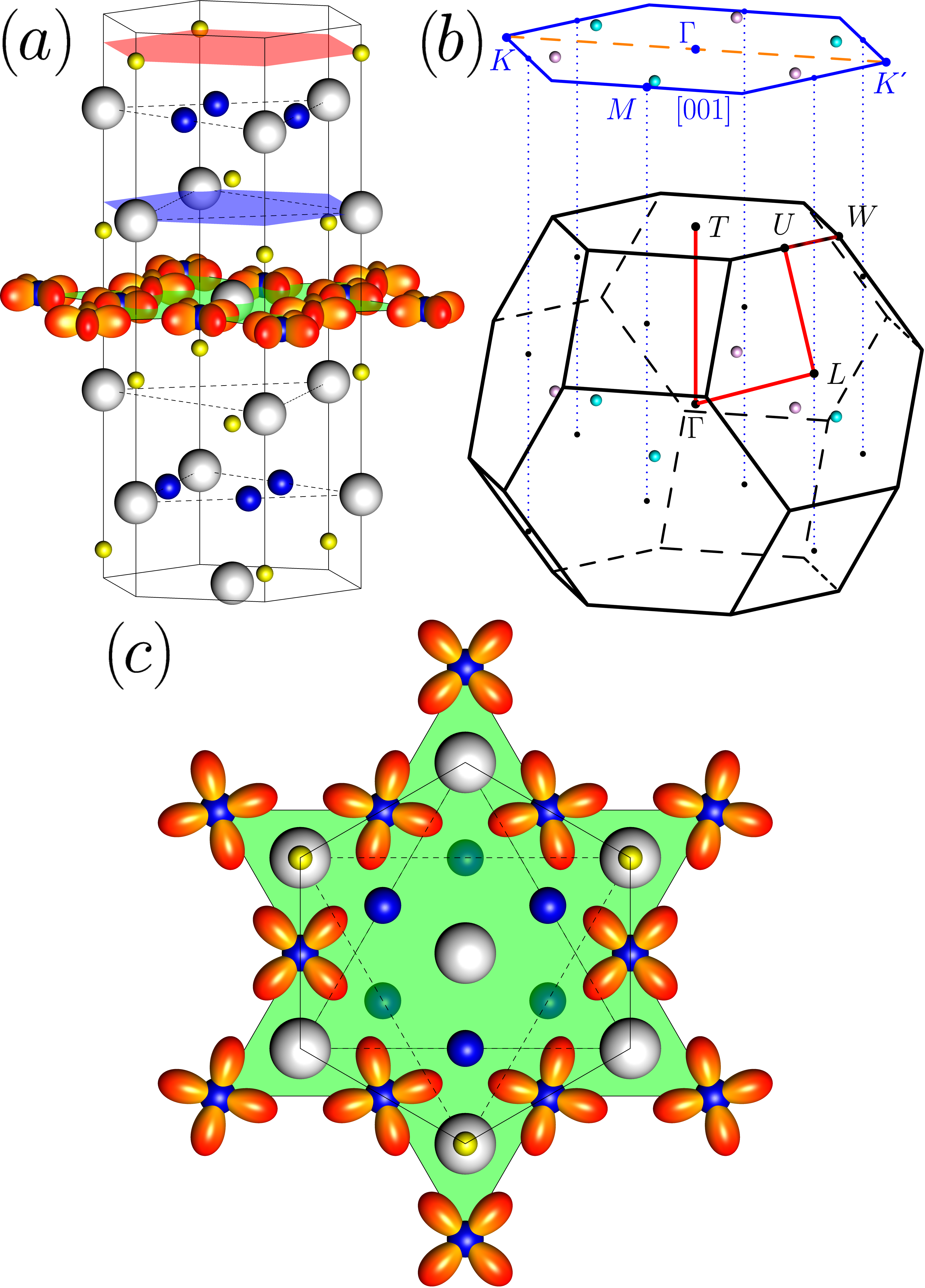}
	\caption{
		Structure and BZ of Co$_3$Sn$_2$S$_2$. (a) Shandite--type crystal structure of Co (blue), Sn (white), and S (yellow) atoms. 
		Colored planes mark the S--type (red), Sn--type (blue) terminations, and Co kagome sublattice (green).
		(b) BZ and projection along [001], with high-symmetry points indicated. Dashed orange line on the two--dimensional projection shows the high--symmetry $\Gamma-K$ cut. Pink/cyan spheres mark Weyl points in the BZ and their surface projections.
		(c) Kagome plane Co-$3d_{x^2-y^2}$ orbitals (red), rotated by $-\pi/12$, $\pi/12$, or $\pi/4$, to align with local $C_3$ cluster symmetry.  
	}
	\label{structure}
\end{figure}

Co$_3$Sn$_2$S$_2$ crystallizes in the hexagonal shandite-type structure \cite{struct1,struct2,synth1,synth2,synth3} with A-B-C stacking in the [001] direction (space group $R\bar{3}m$, \myhash166), shown in Figure \ref{structure}a. The layers alternate as ...-Sn-[S-(Co$_3$-Sn)-S]-... units, with the Sn layer and Co$_3$-Sn layer separated by sulfur atoms. The Co atoms form a Kagome lattice with Sn atoms positioned at the center of each hexagon (Fig. \ref{structure}(c)). 

Magnetic order in this compound has been extensively characterized by experiments \cite{Kubodera2006,Schnelle2013,se-doped-mossbauer,fe-doped,Yin2019}. Below $T_c$, Co$_3$Sn$_2$S$_2$ exists in a ferromagnetic phase, with a local moment of $\sim0.33 \mu_B/$Co atom along the easy magnetization c--axis. Recent experiments have also suggested the existence of an anomalous magnetic phase just below the transition temperature \cite{Kassem2017, Kassem2020,Guguchia2020}. Within the ferromagnetic phase, broken $\mathcal{T}$ symmetry leads to the formation of Weyl points just above the Fermi energy. This band structure topology has been verified through measurements of the spectroscopy \cite{Morali2019, liu_magnetic_2019,jiao_signatures_2019}, anomalous Hall conductivity \cite{Liu2018,Li2020}, anomalous Nernst conductivity \cite{Guin2019}, negative longitudinal magnetoresistance \cite{Liu2018}, and gigantic magneto-optical response \cite{Okamura2020}. 

\begin{figure}[ht]
	\includegraphics[width=\columnwidth]{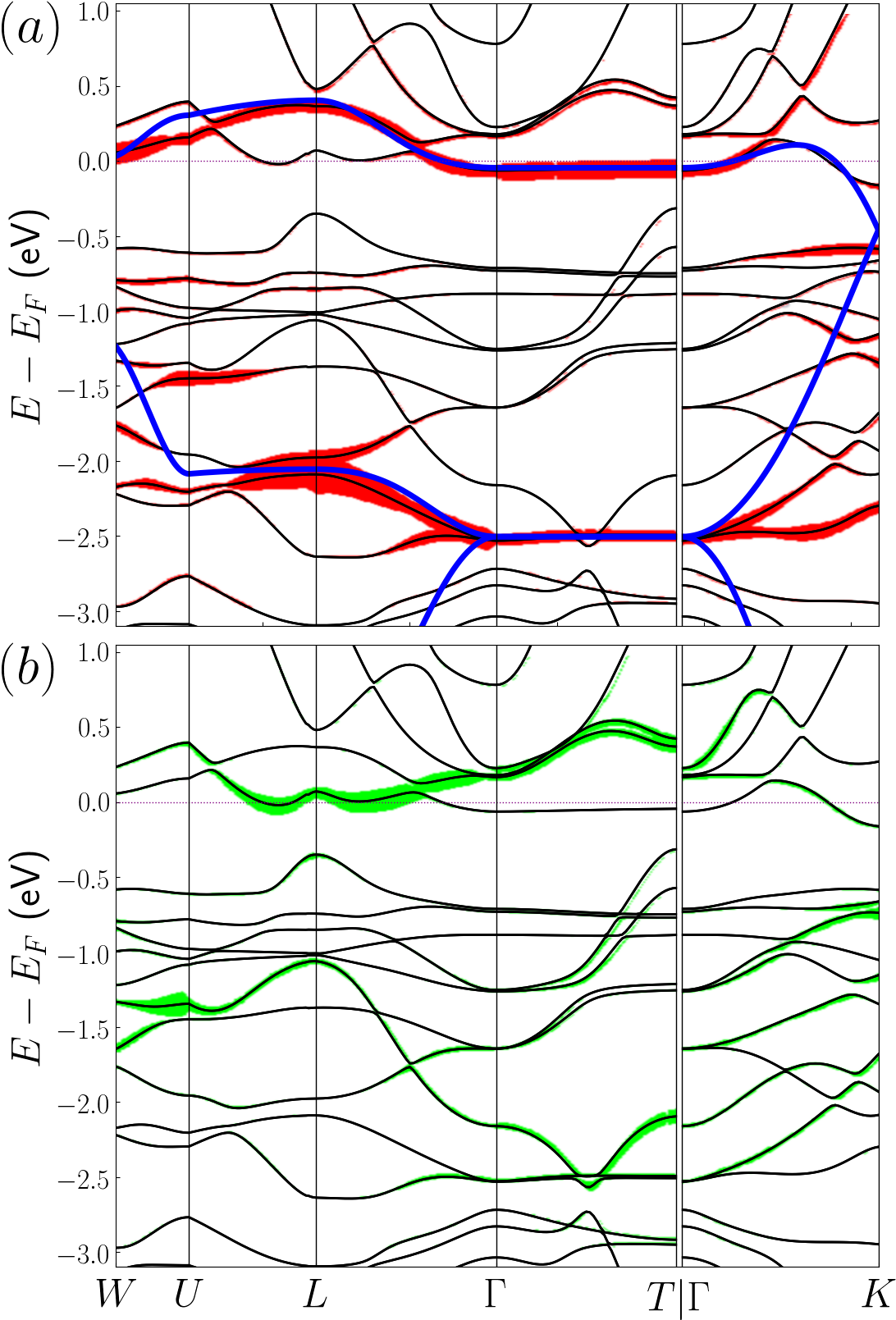}
	\caption{LDA band structure of Co$_3$Sn$_2$S$_2$. The orbital characters for the Co-$3d$ basis, rotated about $k_z$ by $-\pi/12$, $\pi/12$, and $\pi/4$, are highlighted red for $d_{x^2-y^2}$ (a) and green for $d_{z^2}$ (b). The $d_{x^2-y^2}$ tight-binding model bands are shown in blue. }
	\label{ldabands}
\end{figure}

The magnetic order of Co$_3$Sn$_2$S$_2$ is understood from previous electronic structure calculations as follows\cite{Dedkov_2008,Schnelle2013}. The spin-minority channel develops a gap structure, leaving a Fermi surface composed exclusively of spin-majority states, which makes Co$_3$Sn$_2$S$_2$ a half-metallic ferromagnet.
The states at the Fermi level are comprised primarily of Co-$3d$ orbitals, but there is no consensus on the precise magnetic orbital(s) nor mechanism, with various works proposing minimal tight-binding models that assign the magnetism to Co-$d_{z^2}$ \cite{tb-weyls}, Co-$J_{1/2}$ \cite{tb-nodal-lines}, or Co-$3d$-$e_g$ \cite{HalfAP-VI}.
%
%
%
However, the standard approach is to extract a tight-binding model directly from the electronic structure calculation, by projecting onto a basis of maximally localized Wannier functions \cite{Xu_fermiArc_Co3Sn3S2_2018, Morali2019,Yin2019,Wang2018,tb-nodal-lines}. Nevertheless, this procedure is cumbersome and necessitates an excess of orbital fitting parameters in order to adequately reproduce the electronic bands. The large number of Wannier orbitals needed for most fits, and the possibility of fitting bands with the wrong orbitals, makes it difficult to analyze the electronic structure and physics of complex materials.

To understand the origin of the magnetism in Co$_3$Sn$_2$S$_2$, we employ the full--potential linear muffin--tin orbital (FPLMTO) method \cite{FPLMTO,sav1992} to perform density functional theory calculations. The electronic structure is computed within the local density approximation (LDA), local spin density approximation (LSDA), and with the Hubbard-$U$ correction included (LDA+U) to take into account the electron--electron interactions on the Co-$3d$ orbitals. In order to best match experimental results, the parameter $U$ is varied from $U = 2-4$ eV and with Hund's $J_H = 0.8$, in accordance with recent many--body simulations \cite{NatComm-DMFT}.

The LDA electronic structure calculation reveals two bands crossing the Fermi energy (Fig. \ref{ldabands}), carrying mainly Co-$3d$ character. To obtain a clearer picture, we project onto a basis set of cubic harmonics that are rotated about the $k_z$ axis by $-\pi/12$ for Co-1, $+\pi/12$ for Co-2, and $+\pi/4$ for Co-3, which respects the local $C_3$ rotational symmetry and aligns the orbitals with the in--plane nearest--neighbor Co bonds (Fig. \ref{structure}(c)). In this new basis, the two bands crossing the Fermi level have $d_{x^2-y^2}$ character for the half-filled flat band that dips below $E_F$ along $\Gamma-T$, and $d_{z^2}$ character for the other, mostly unoccupied band. Below $E_F$, the dispersion of local $d_{x^2-y^2}$ orbitals (red fat lines in Fig. \ref{ldabands}(a)) is more complicated due to hybridization with many other bands, which can be seen within a large energy window from -2.5 eV to -0.5 eV.

The lack of dispersion along $\Gamma-T$ for the Co-$d_{x^2-y^2}$ band implies that the electron hoppings are confined within the kagome plane. It is well known that a tight-binding model restricted to only include nearest-neighbor in--plane hoppings produces an ideally flat band \cite{Yin2019}. For the case of the band structure here, we see some additional dispersion along $U-L-\Gamma$, which suggests higher order hoppings. Indeed the band structure can be closely reproduced (Fig. \ref{ldabands}) by including couplings up to third order \cite{supplement}, with on--site energy $\epsilon_{\text{Co}} = -1.87$ eV, and hoppings $t_1 = 0.77$ eV, $t_2 = -0.36$ eV, and $t_3 = 0.046$ eV. These parameters are similar to those used for tight-binding models in previous works \cite{tb-weyls, tb-nodal-lines}, notwithstanding the different basis sets that were employed.

\begin{figure}[ht]
	\includegraphics[width=1.0\columnwidth]{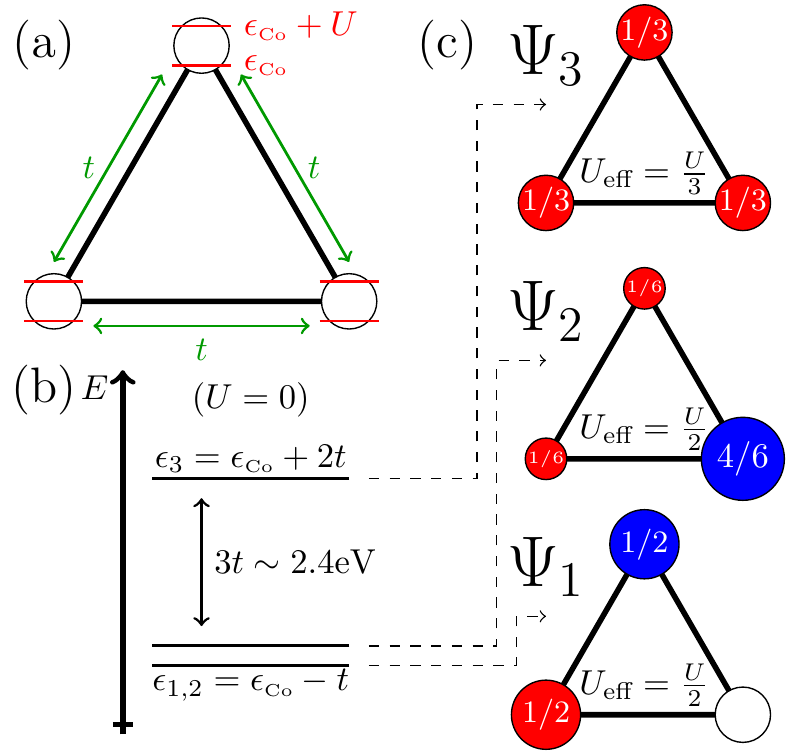}
	\caption{
		(a) Local Hubbard model on the three Co atom cluster. Diagonalizing the non-interacting case yields a pair of degenerate states and one higher energy state (b), with the probability density distributions shown in (c). Coulomb interactions are screened for $\Psi_3$, yielding $U_{\text{eff}} = U/3$.
	}
	\label{cluster}
\end{figure}

The basic structure of the energy bands can be illustrated by considering what happens locally on a triangular Co atom cluster (Fig. \ref{cluster}), with $\psi_{\text{Co}1}, \psi_{\text{Co}2}, \psi_{\text{Co}3}$ forming the basis of orbitals on the three Co atoms. Considering only onsite and nearest--neighbor hoppings in the non-interacting picture yields the $3\times 3$ Hamiltonian
\begin{equation}
	\mathcal{H} = 
	\begin{pmatrix}
	\epsilon_{\text{Co}} & t_1 & t_1 \\
	t_1 & \epsilon_{\text{Co}} & t_1 \\
	t_1 & t_1 & \epsilon_{\text{Co}} \\
	\end{pmatrix},
\end{equation}
which can be diagonalized to find the eigenstates $\Psi_1 = (-\psi_{\text{Co}1}+\psi_{\text{Co}2})/\sqrt{2}$, $\Psi_2 = (\psi_{\text{Co}1}+\psi_{\text{Co}2}-2 \psi_{\text{Co}3})/\sqrt{6}$, and $\Psi_3 = (\psi_{\text{Co}1}+\psi_{\text{Co}2}+ \psi_{\text{Co}3})/\sqrt{3}$. $\Psi_{1,2}$ form a doubly degenerate ground state with energy $\epsilon_{1,2} = \epsilon_{\text{Co}} - t_1$, lying below $\Psi_3$ which has an energy $\epsilon_{3}=\epsilon_{\text{Co}} + 2t_1$.

We now diagonalize this model with an onsite Coulomb repulsion $U$, and consider what happens for different electron fillings. At half-filling the two lowest energy configurations are: $S=1/2$ with all three electrons in the lower energy doublet, and the higher energy $S=3/2$ where one electron occupies $\Psi_3$. Adding an additional electron preferentially fills the hole in the doublet, resulting in a $S=0$ ground state and $S=0,1$ excited configuration. At a filling of 5/6 electrons the situation is simplified; the lowest energy configuration matches the single electron picture with the solitary hole occupying $\Psi_3$, which corresponds to $S=1/2$. 

This in fact matches what is seen in LDA calculations; the highest--lying $d_{x^2-y^2}$ band is half filled, in line with a five electron/one hole configuration. This results in a spin-$1/2$ ferromagnetic state, since the antiferromagnetic state is impeded by lattice frustration. The associated moment of 1 $\mu_B$ is spread over the Co atoms in the $\Psi_3$ state, consistent with the 0.33$\mu_B$ per Co magnetic moment computed by LSDA. The experimentally measured moment is slightly less than 1 $\mu_B$/f.u. \cite{Schnelle2013, se-doped-mossbauer, supplement}, due to the partial filling of the $d_{z^2}$ band.

The $\Psi_3$ state is optimal from the standpoint of minimizing Coulomb repulsion. For $\Psi_1$ and $\Psi_2$ the Coulomb repulsion is slightly reduced to $U/2$, while the symmetric  distribution for $\Psi_3$ results in a net repulsion of $U/3$. Considering a range $U\sim 2-4$eV, which is typical for Co atoms in magnetic systems, the electron filling up to the half--filled $\Psi_3$ state reduces this to values comparable with the bandwidth (Fig. \ref{ldabands}) found by first--principles calculations \cite{supplement}. This means that the disordered magnetic phase near $T_c$ (to be considered in a later section), is a true strongly correlated state, and must be treated appropriately. Recent single--impurity DMFT simulations were not able to fully capture the electronic-bandwidth renormalization \cite{NatComm-DMFT}. Thus a complete treatment would entail a calculation that captures the full physics of the three-Co-atom clusters, through a method such as cluster-DMFT \cite{Haule-cluster-DMFT}. Such an approach is outside the scope of the present study, but may yield important insights in the future.

\section{III. Surface Terminations.}

\begin{figure}[b]
	\includegraphics[width=1\columnwidth]{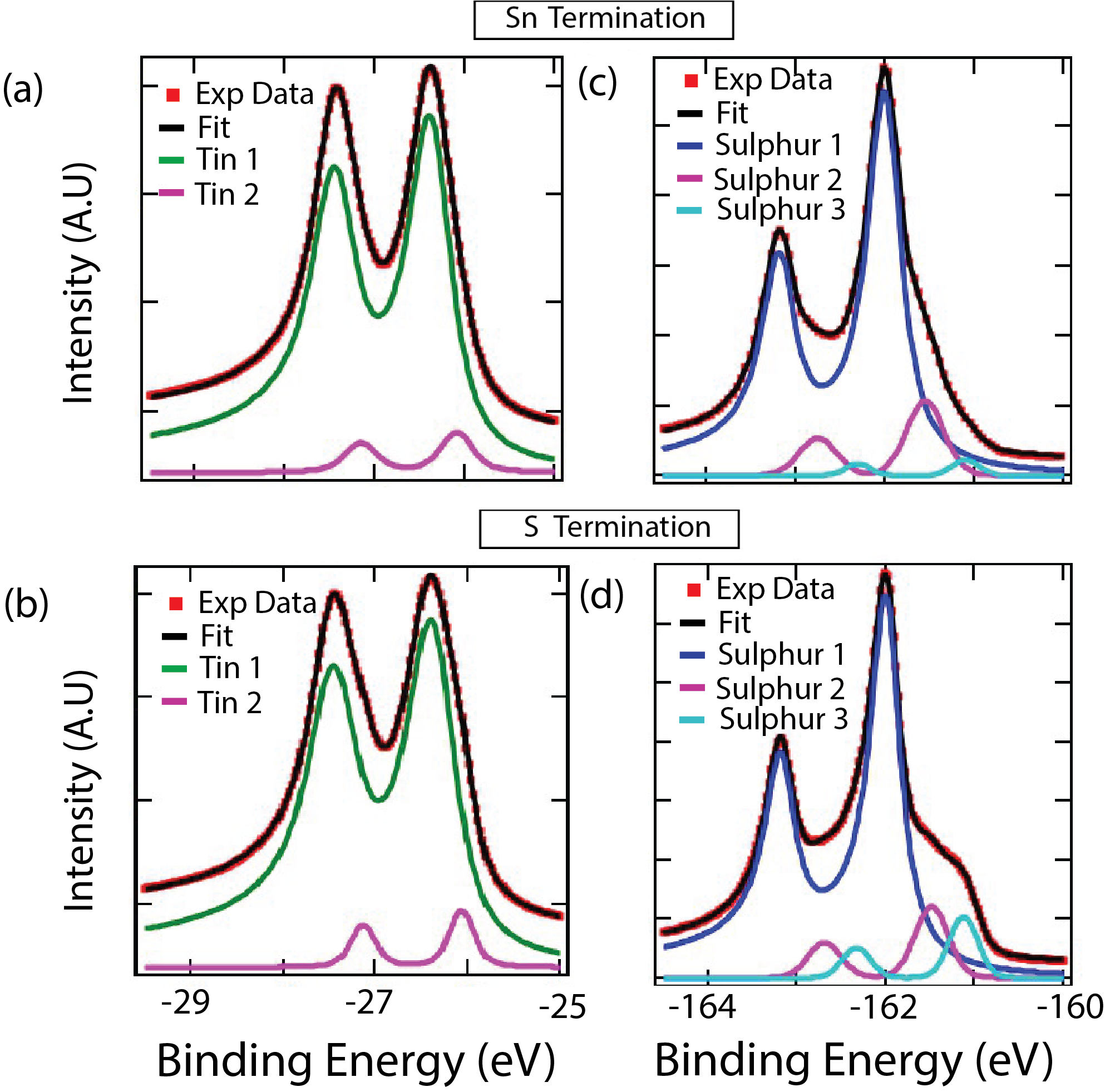}
	\caption{
		Sn and S core levels on different surface terminations. (a)-(b) Sn 4d core level spectra taken on portions of the sample with Sn (a) and S (b) terminations. Core levels are fitted with two DS doublets in order to represent the contribution from the first two Sn layers. (c)-(d) S 2p core level spectra at the same positions of the sample with Sn(c) and S(d) terminations fitted with three DS doublets representing the contribution from the first three S layers.  All spectra taken at 30K.
	}
	\label{fig:xps}
\end{figure}

From a theory standpoint, the presence of Weyl points in the bulk guarantees the existence of topological Fermi arcs on the surface \cite{WSMRMP,Savrasov_arcs}, but the shape and connectivity of these arcs have been shown to critically depend on surface termination \cite{Rossi_PRB}.
Co$_3$Sn$_2$S$_2$ cleaves by breaking either Co--S or Sn--S bonds, with the latter being more probable \cite{Xu_fermiArc_Co3Sn3S2_2018}, making Sn and S the dominant terminations. These cleave terminations in Co$_3$Sn$_2$S$_2$ have implications for Fermi arc connectivity, resulting in \textit{intra}--BZ Fermi arcs on the Sn termination, \textit{inter}--BZ Fermi arcs on the Co termination \cite{Morali2019}, and very short Fermi arcs on the S-termination due to mixing with bulk states \cite{Xu_fermiArc_Co3Sn3S2_2018}.  Additionally, the S-termination is susceptible to vacancy formation, and these vacancies themselves yield novel excitations \cite{Xing_SpinOrbitPolaron_Co3Sn2S2_2020, Gao2021_defects}.  As such, identifying the photoemission fingerprints of different terminations is crucial for the study of temperature--dependent band structure and topology.


Photoemission experiments were performed at beamline 7.0.2 (MAESTRO) at the Advanced Light Source (ALS), with a beam spot size $d\approx$ 80$\mu$m.  115eV photon energy was used for ARPES measurements, which accesses near the $\Gamma$ plane of the BZ and has been shown to cut through Weyl points \cite{liu_magnetic_2019}. X-ray photoemission spectroscopy (XPS), with 350eV photon energy was used to assess core levels at each measured sample position. 

\begin{figure}[htb]
	\includegraphics[width=1.0\columnwidth]{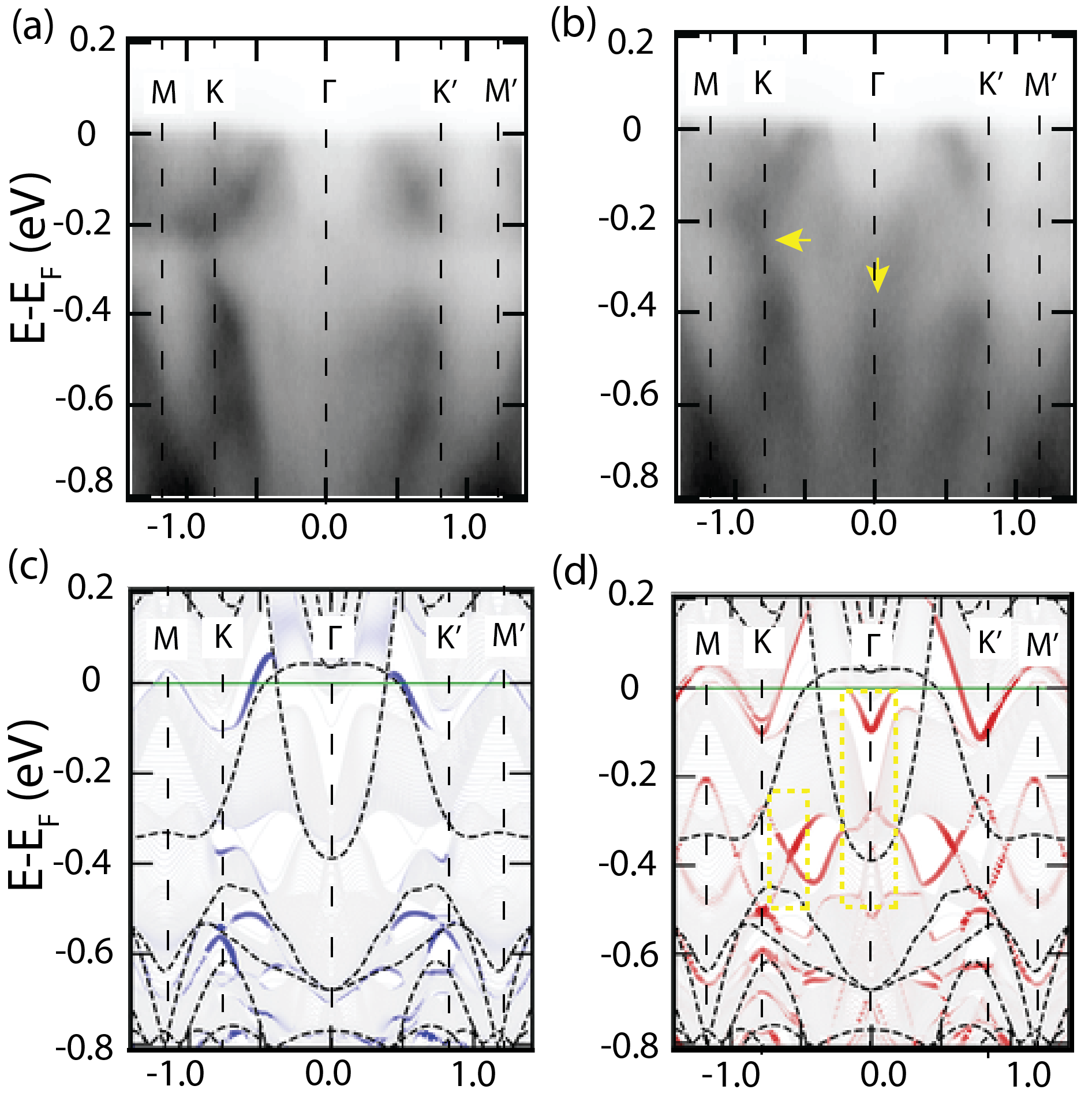}
	\caption{ High symmetry $\Gamma-K$ cuts taken at 30K for Sn (a) and S (b) terminations.
	Arrows point to features near $\Gamma$ and K which differ between the two terminations. (c)-(d) Slab calculations along $M-K-\Gamma-K^\prime-M^\prime$. Dashed lines indicate bulk bands. Slab bands are colored based on surface character: blue--Sn termination, red--S termination, grey--bulk. Vertical dashed lines mark high symmetry points throughout.}
	\label{fig:fig1}
\end{figure}

Figures \ref{fig:xps} and \ref{fig:fig1} show results for the two surface terminations realized at different positions in our Co$_3$Sn$_2$S$_2$ samples. These are identified as Sn and S surfaces, respectively indicated by the blue and red planes in Fig. \ref{structure}. The most probable identity of the termination in different positions on the same sample is determined from density functional theory (DFT) calculations, which are presented later. These different terminations are most readily distinguishable by their S $2 p_{1/2,3/2}$ core level spectra shown in Figure \ref{fig:xps} (a)-(b), which display an extra low-energy shoulder on the S termination  (Fig. \ref{fig:xps}(b)).  Sn $4 d_{3/2,5/2}$ core levels were also measured (\ref{fig:xps} (c)-(d)), and do not show as visually apparent differences as a function of termination.  XPS spectra were fitted to Doniach-Sunjic (DS) doublet profiles with a Shirley background. Three sets of doublets are needed to fully capture the S core levels: one associated with bulk states and two associated with two near-surface sulphur layers.  The latter contribute more strongly on the S termination, producing the characteristic shoulder.  Sn core levels were fitted to two DS doublets, but have only subtle lineshape differences at different cleave terminations.

\begin{figure*}[htb]
	\includegraphics[width=\textwidth]{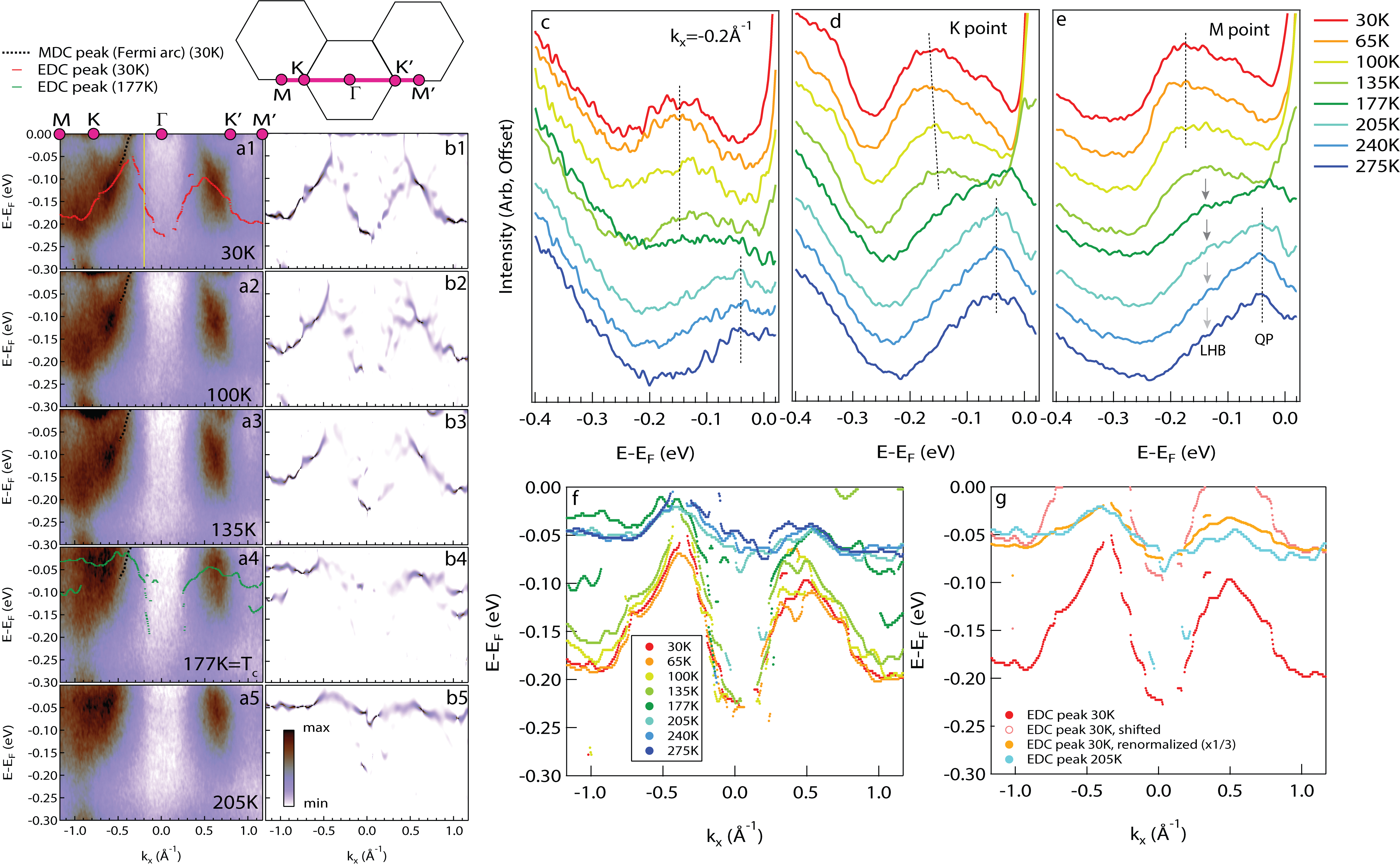}
	\caption{Temperature dependence of the $\Gamma-K$ cut. (a) Selected temperatures along cut indicated in schematic above panels.  In all spectra, a Fermi-Dirac cutoff convolved with the energy resolution has been divided out.  Black dashed line indicates position of band which forms Fermi arc, as quantified by MDC fitting at 30K.  The same dispersion is plotted in  a1-a4. b) corresponding curvature plots \cite{Zhang:curvaturePlot} at each temperature c)-e) temperature-dependent EDCs at k$_x=$0.2, K point, and M point, respectively.  Curves normalized at $E_B=1$ eV and offset for clarity.  Dashed lines are guide-to the eye for EDC peak position in each panel, and grey arrows in e) are guide-to-the-eye for high-energy shoulder. LHB and QP label the lower hubbard and quasiparticle bands, as discussed in text. All EDCs have Fermi-Dirac function divided out. f) EDC local maxima at each studied temperature. g) comparison of band positions as quantified by EDC local maxima at 30K ($<$Tc) and 205K ($>$Tc).  A rigid band shift by 130meV is contrasted with renormalization by 1/3 for comparison with 205K data}
	\label{fig:TdepGK}
\end{figure*}

Having established the presence of two distinct surface terminations, we now turn to the measured and calculated band structure at the same positions as the XPS spectra. Fig. \ref{fig:fig1} (a)-(b) shows ARPES spectra along the $\Gamma - K$ high symmetry direction for Sn (a) and S (b) terminations.  The cut extends into the second BZ, and high symmetry points are labeled. Low energy features which are most relevant to magnetism and topology along this high-symmetry  cut are discussed in Fig. \ref{fig:TdepGK} and SM \cite{supplement}. Here, we focus on characteristic differences between the two cleave terminations. The most prominent spectral changes appear at the $\Gamma$ point (vertical yellow arrow in \ref{fig:fig1} (b)) and also at $k_x\approx 0.8 $\AA$^{-1}$ near the $K$ point (horizontal yellow arrow in \ref{fig:fig1} (b)). 
At the $\Gamma$ point, a `Y-shaped' feature is prominent on the S termination (\ref{fig:fig1}(b)), and significantly diminished on the Sn termination (\ref{fig:fig1}(a)). 
Meanwhile, on the Sn termination, there is a gap or depletion of spectral weight near the K point, at a binding energy $E_B \sim 300$ meV, which is more filled in on the S termination. 


In order to interpret the position-dependent ARPES spectra, we extract a model Hamiltonian from the \emph{ab initio} calculation to solve on a layered slab geometry using the following procedure. We use the Quantum {ESPRESSO} package \cite{QE1,QE2} for first--principles electronic structure calculations of Co$_3$Sn$_2$S$_2$ within LDA and LSDA, which are performed using projector augmented waves (PAW) in the plane wave basis with spin-orbit coupling included. Scalar relativistic PAW pseudopotentials are employed, with exchange and correlation terms included through the Perdew-Burke-Ernzerhof (PBE) parameterization scheme as implemented in pslibrary21 \cite{PSL}. The self-consistent calculation is performed using the experimental lattice constants \cite{Vaqueiro2009}, on a grid of $10\times10\times10~ \bm{k}$-points with an energy cutoff of 60 Ry, and subsequently projected onto maximally localized Wannier functions using the {WANNIER90} package \cite{W90}, with a starting basis of Co-3\emph{d}, Sn-5\emph{sp}, and S-3\emph{p} orbitals. 

The resulting tight-binding Hamiltonian in the Wannier function basis is used to construct a 60 unit cell slab for the surface state calculations. The supercell Hamiltonian is assembled along the [001] crystallographic direction and truncated along the Sn and S surfaces shown in Figure \ref{structure}. To isolate the surface states for comparison with ARPES spectra, we color the bands red(blue) at each $k$-point proportionally to the probability density at the S (Sn) surface for their corresponding eigenvectors (Fig. \ref{fig:fig1}(c)-(d)).  Several additional layers of atoms are included in our definition of the slab surface, to account for the long penetration depth of the surface states into the bulk \cite{Xu_fermiArc_Co3Sn3S2_2018}, and to incorporate the contribution of the Co atoms, since the surface states for these terminations have primarily Co-$3d$ character  \cite{Li_oxidation_2019}.

Numerical slab calculations of the S termination show numerous surface bands at the $\Gamma$ point, which do not appear on the Sn terminated surface, motivating us to identify the spectrum in Fig. \ref{fig:fig1}(b) with the S termination.  Additionally, the calculation for the S surface shows a band in the gap between the two bulk states at $K$, which is also not present in the Sn termination.  This is consistent with the filled--in spectral weight at $E_B \sim 300$ meV in Fig. \ref{fig:fig1}(b) as compared to (a).  This interpretation is also consistent with prior XPS studies, which associate the low-energy shoulder on the S-2p levels with surface S atoms, whose energy shift stems from reduced coordination at the surface \cite{Holder_PhotoemissionCo3Sn2S2_2009, Li_oxidation_2019}.  The stronger shoulder in Fig. \ref{fig:xps}(b) vs (a) thus suggests a greater contribution from surface S atoms, and a predominant S termination on the sample position used for \ref{fig:xps}(b) and \ref{fig:fig1}(b).  We note that the spectra associated with different cleaves were observed with 80 $\mu$m photoemission spot size, which suggests that different cleave terminations typically have macroscopic dimensions, and in fact some of our experiments yielded only a single termination.  We also acknowledge studies of nanoscale inhomogeneity within that macroscopic cleave termination \cite{Morali2019,roccapriore2021revealing}, which our simplified model of a single termination does not consider.

\section{IV. Electronic Structure Changes Across T$_c$.}


To describe the physics of the magnetic transition and associated changes in the electronic structure, we present ARPES measurements of Co$_3$Sn$_2$S$_2$ across $T_c$, focusing on the high symmetry cut along $\Gamma-K$ in the 2D projected BZ shown in Fig. \ref{structure}(b).


Fig. \ref{fig:TdepGK} investigates the temperature dependence of low-energy electronic structure through $T_c$ on the Sn termination.  Example spectra are shown at 30K ($T<<T_c$), 100K, 135K, 177K ($T=T_c$), and 205K ($T>>T_c$) in panels a1-a5.  Corresponding curvature plots \cite{Zhang:curvaturePlot} (Fig. \ref{fig:TdepGK} b) highlight less visible features in the spectra, particularly the parabolic dispersion around $\Gamma$. Although this feature has lower cross section than other bands, it still yields a clear local maximum in the EDC (Fig. \ref{fig:TdepGK}(c) and SM \cite{supplement}).

In the energy and momentum range shown in Fig. \ref{fig:TdepGK} the predominant features observed are bulk bands, highlighted in red in panel a1, and a surface-like band which forms the Fermi arc, highlighted with a black dashed line. The bulk and surface character of these respective features was established in prior photon-energy dependence studies \cite{liu_magnetic_2019}. These features are in qualitative agreement with calculations shown in Fig. \ref{fig:fig1}, as discussed further in SM \cite{supplement}.  The band which forms the Fermi arc is largely unchanged at 100K, but at 135K, this feature is weakened and shifted towards the K point.  At 177K, it is no longer present.

The bulk bands are best visualized via the EDC local maximum, which also corresponds well with the dominant features in the curvature plots.  Fig. \ref{fig:TdepGK}(c)-(e) show EDCs at three characteristic momenta \textemdash~ $k_x=$0.2 $\AA^{-1}$, K point, and M point \textemdash~ to emphasize the robustness of the observed energy shifts at each momentum.  Additionally, EDCs at the M point show clear high-energy shoulders above T$_c$, indicating persisting features of the magnetic state, identified as the lower Hubbard band.

Fig \ref{fig:TdepGK}(f) shows EDC local maxima across the cut.  The dispersions derived in this manner coincide at all measured temperatures below T$_c$, but are distinct from those measured above T$_c$. Intermediate behavior is shown at 177K$=$T$_c$, both in the EDC local maxima and in the curvature plots. Above T$_c$, EDC local maxima are consistently shifted to lower binding energy, but the shift is non-monotonic: at the $\Gamma$ point, the shift is 150 meV,  at the $K$ point the shift is 70 meV, and at the $M$ point the shift is 130 meV.  Dispersions above $T_c$ are significantly flatter than below T$_c$, indicating an enhanced effective mass relative to the ferromagnetic regime.  This is demonstrated in Fig. \ref{fig:TdepGK}(g) where a characteristic dispersion T$<$T$_c$ is compared to a characteristic dispersion T$>$T$_c$ in two different ways: energy shift and renormalization.  A simple shift does not reproduce the $T>T_c$ dispersion, but a renormalization by a factor of $\sim 3$ produces a dispersion coincident with the one measured above T$_c$.

\begin{figure*}[htb]
	\includegraphics[width=\textwidth]{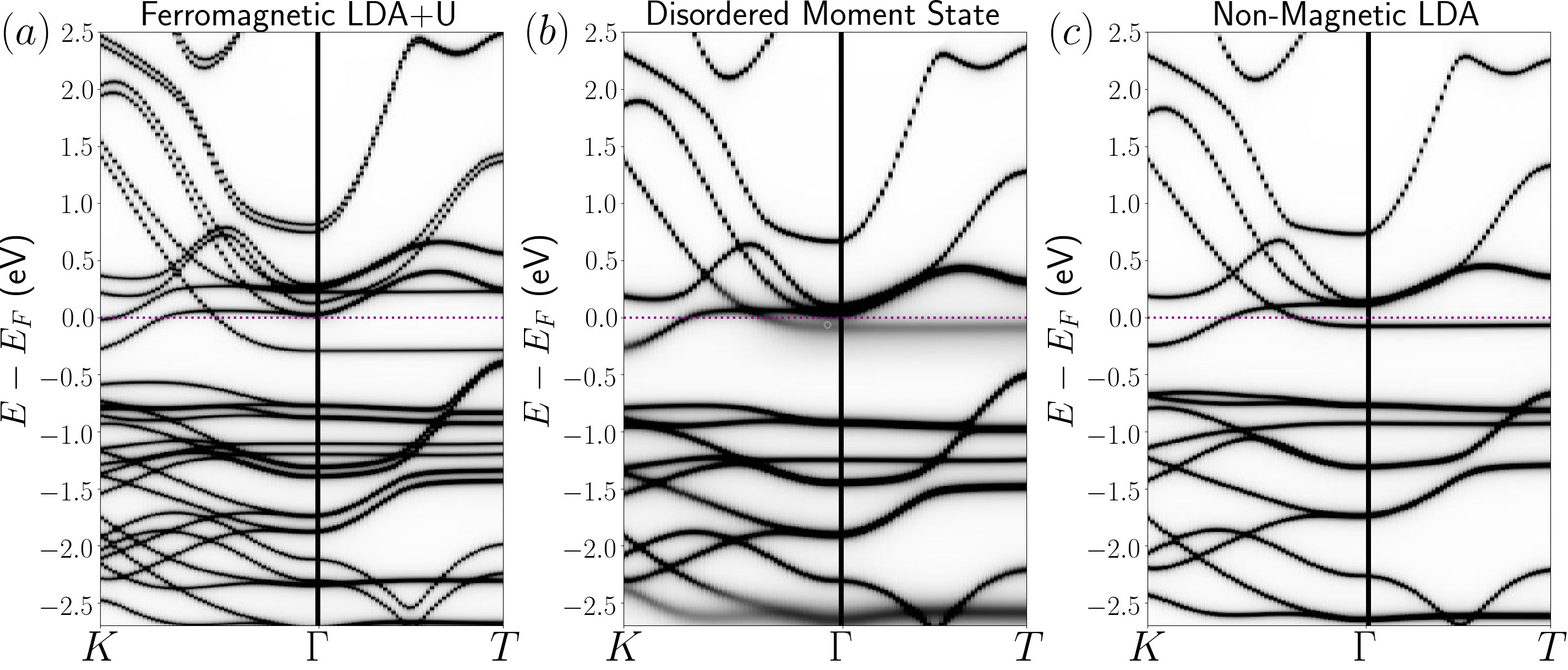}
	\caption{Plots of the momentum--resolved spectral function $\mathcal{A}(\bm{k},\omega) = \Im [G(\bm{k},\omega)]$ along $K-\Gamma-T$. The spectral functions for (a) ferromagnetic LDA+U, (b) LDA+DLM, and (c) non--magnetic LDA are shown.
	}
	\label{dlm}
\end{figure*}

As we have already discussed, the ferromagnetism of Co$_3$Sn$_2$S$_2$ below $T_c$ serves to split the electronic bands within the spin--majority and spin--minority channels \cite{supplement}, resulting in a band gap in the spin-minority states. One might naively attribute the apparent shift in the band structure observed by experiments solely to the disappearance of exchange splitting in this itinerant magnet. If the magnetic moments are presumed to disappear entirely, the state where exchange splitting vanishes could be represented using the electronic structure predicted by non-magnetic LDA. Then, the behavior across the ferromagnetic--paramagnetic transition can be identified with the shift of the spin--majority band between magnetic LSDA and non-magnetic LDA \cite{yang_magnetization-induced_2020, Holder_PhotoemissionCo3Sn2S2_2009}.


While it is tempting to compare the results of the non--magnetic calculation with the behavior of Co$_3$Sn$_2$S$_2$ above $T_c$ in this way, such an interpretation does not entirely capture the physics of the paramagnetic phase. The magnetism is only fully quenched at temperatures comparable to the band splitting of $\sim 0.5$ eV, far above the melting point of the material. For experimentally relevant temperatures above $T_c$, ferromagnetism disappears due to the suppression of long--range order by temperature driven fluctuations. However, the persistence of local moments seen in the Curie Weiss susceptibility \cite{supplement}, suggests a more complex many--body behavior.

The microscopic picture of magnetism is usually considered in the limit of large, moderate, and small ratios of Coulomb interaction $U$ to the electronic bandwidth $W$. When $U>>W$, we are in the archetypal Mott regime where disorder can narrow the bands by further limiting inter--site hopping, while $U<<W$ corresponds to the standard itinerant state. However, when $U\sim W$, we recover a distinctive strongly correlated state, 
where renormalized quasiparticles in the vicinity of $E_F$ and Hubbard band states at higher energies are simultaneously present.

Applying this interpretation of magnetic disorder to Co$_3$Sn$_2$S$_2$, it becomes clear that that the relative magnitudes of the bandwidth and band splitting will determine the behavior above $T_c$. For a large band splitting, magnetic disorder and associated correlations will simply serve to renormalize the width of the Hubbard bands. On the other hand, if $U\sim W$, the magnetic disorder at high $T$ will lead to a narrow quasiparticle band, resulting in a correlated metallic state, appearing as a spectral shift in photoemission measurements \cite{Kotliar-spectral}.

The expected behavior can be estimated from the parameters of our extracted tight--binding model. The nearest--neighbor hopping parameter is approximately $t \sim 0.8$eV, while a range of Hubbard--$U$ values $U \sim 2-4$eV result in a reasonable exchange splitting between the bands. For the Co-$3d_{x^2-y^2}$ three band model at 5/6 filling, the highest occupied state (Fig. \ref{cluster}) renormalizes the Coulomb repulsion by a factor of three. 
Representing this system as a simplified single--band Hubbard model with a bandwidth $W \sim 0.8$eV and effective $U_{\text{eff}}\sim 1.3$eV, places it in the regime $U \sim 1.6 W$ which is known to be very close to a Mott transition \cite{Kotliar1996}, although multi--orbital effects should affect the precise $U/W$ ratio \cite{Kotliar-mott}. Thus, it is expected that Co$_3$Sn$_2$S$_2$ would transition to a correlated metallic state once the moments are disordered by temperature. 

There are numerous signatures of correlations in Co$_3$Sn$_2$S$_2$, but a complete picture is only beginning to emerge. Even in the ferromagnetic phase there is a modest enhancement of the electronic mass, with the experimental Sommerfeld gamma $\gamma_{\text{expt}} = $ 9.8 -- 10.8  mJ mol$^{-1}$ K$^{-1}$ \cite{kassem_2017_thesis, Schnelle2013} exceeding the value computed from the electronic density of states $\gamma_{\text{th}} = 3.25$ mJ mol$^{-1}$ K$^{-1}$ \cite{supplement} by a factor of $\sim 3$. However, it should be noted that mass enhancement will be unequal across the bands, with correlation effects playing a more significant role in the renormalization of the magnetic Co-$3d_{x^2-y^2}$ band. In fact, the electronic-bandwidth renormalizations of Co-$3d_{xy}$ and Co-$3d_{x^2-y^2}$ are found to be strongest with increasing Coulomb energy \cite{NatComm-DMFT}, which is consistent with our description, as these are the orbitals that most align with the rotated Co-$3d_{x^2-y^2}$ orbitals in our symmetry--adapted basis. Further evidence for correlations comes from optical conductivity measurements which show a moderate band renormalization that necessitates a rescaling of the computed spectrum by a factor of $\sim$ 0.68 -- 0.76 to match experiment \cite{NatComm-DMFT, yang_magnetization-induced_2020}.  This is consistent with the recently measured ratio of 0.70 between experimental and theoretical bandwidths \cite{liu_magnetic_2019}.

This wide assortment of correlated properties is further enriched by temperature and doping effects. A frustrated in-plane anti-ferromagnetic component emerges at a temperature $T_a \sim 90$ K, that tilts the Co magnetic moments \cite{Liu2018, Guguchia2020}, affecting the long-range correlations and resulting in a non-monotonic magnetoelastic response for temperatures between $T_a$ and $T_c$ \cite{Liu2021_magnetoelastic}. As the temperature is raised through $T_c$, optical conductivity measurements also show a transfer of spectral weight to higher energy peaks \cite{yang_magnetization-induced_2020}, suggesting a complex magnetic transition. Doping with electrons or holes reveals additional anomalous behaviors. For instance hole doped Co$_3$Sn$_{2-x}$In$_x$S$_2$ displays a strong enhancement of the quadratic term in the electrical resistivity $\rho(T) = \rho_0 + A T^2$ near $x=0.8$, implying an enhanced electronic mass due to the increased correlation effects \cite{kassem_2017_thesis}. Hole doping with iron leads to an improvement in the thermoelectric properties, which is associated with an enhanced density of states, while electron doping with nickel significantly decreases the resistivity, resulting in a more metallic state with increasing nickel content \cite{thermo-doped}.

A moderate band renormalization by a factor of $\sim 0.7$ would imply that most of the spectral weight resides not in the Hubbard bands, but in the quasiparticle band, which is consistent with the picture that emerges from our ARPES data shown in Fig. \ref{fig:TdepGK}. However, a complete treatment of these effects would involve taking into account the three-Co cluster in a full--fledged many--body calculation \cite{NatComm-DMFT}. Nevertheless, the magnetic transition from ferromagnet to correlated metal in Co$_3$Sn$_2$S$_2$, and the apparent shift of low energy bands seen in photoemission, can be understood using a DLM model \cite{Pindor-DLM}. We compute the electronic structure using LDA+DLM, as well as  in the ferromagnetic LDA+U and non-magnetic LDA settings. The Hubbard $U$ parameter is selected semi--empirically by comparing the position of the flat $d_{x^2-y^2}$ band along $\Gamma-T$. A value of $U=2$eV places this band $0.3$eV below the Fermi level, which is consistent with photoemission measurements \cite{Holder_PhotoemissionCo3Sn2S2_2009}. Alternatively, one can use the larger value of $U$ and adjust the double counting to match with ARPES. The spectral functions $\mathcal{A}(\bm{k},\omega) = \Im [G(\bm{k},\omega)]$ for each calculation are shown in Figure \ref{dlm}.


A comparison of the ferromagnetic and DLM calculations confirms the existence of a correlated regime, clearly showing the exchange split bands collapsing into one, with the DLM result nearly matching the dispersion of the Co-$d_{x^2-y^2}$ band in the non--magnetic calculation. This is an interesting result because the DLM calculation would naively be expected to produce only Hubbard bands and should not differ much from the magnetically ordered LDA+U spectrum. However, the important effect of the DMFT self-consistency included in the Coherent Potential Approximation (CPA), results in the appearance of a quasiparticle band which is seen to be significantly broadened due to quasiparticle life time effects.
Experimentally, this would present as an apparent shift of spectral weight closer to $E_F$, exactly as seen by ARPES measurements.
%
%
We stress that despite the visual similarity between the LDA+DLM and non--magnetic LDA spectral functions, the paramagnetic state we model with DLMs is indeed a complex strongly correlated many-body state, and 
whose band narrowing effect cannot be fully captured by our CPA--based DLM method, nor by single-site DMFT calculations as reported earlier \cite{NatComm-DMFT}.

\section{V. Weyl physics Across T$_c$.}


As the temperature is raised through $T_c$, besides an effective shift in the observed spectral density, the vanishing ferromagnetism also removes the global $\mathcal{T}$--breaking symmetry. Above $T_c$ the system has both $\mathcal{I}$ and $\mathcal{T}$ symmetries, meaning any Weyl points in the ferromagnetic state can no longer exist.

\begin{figure}[ht]
	\includegraphics[width=1\columnwidth]{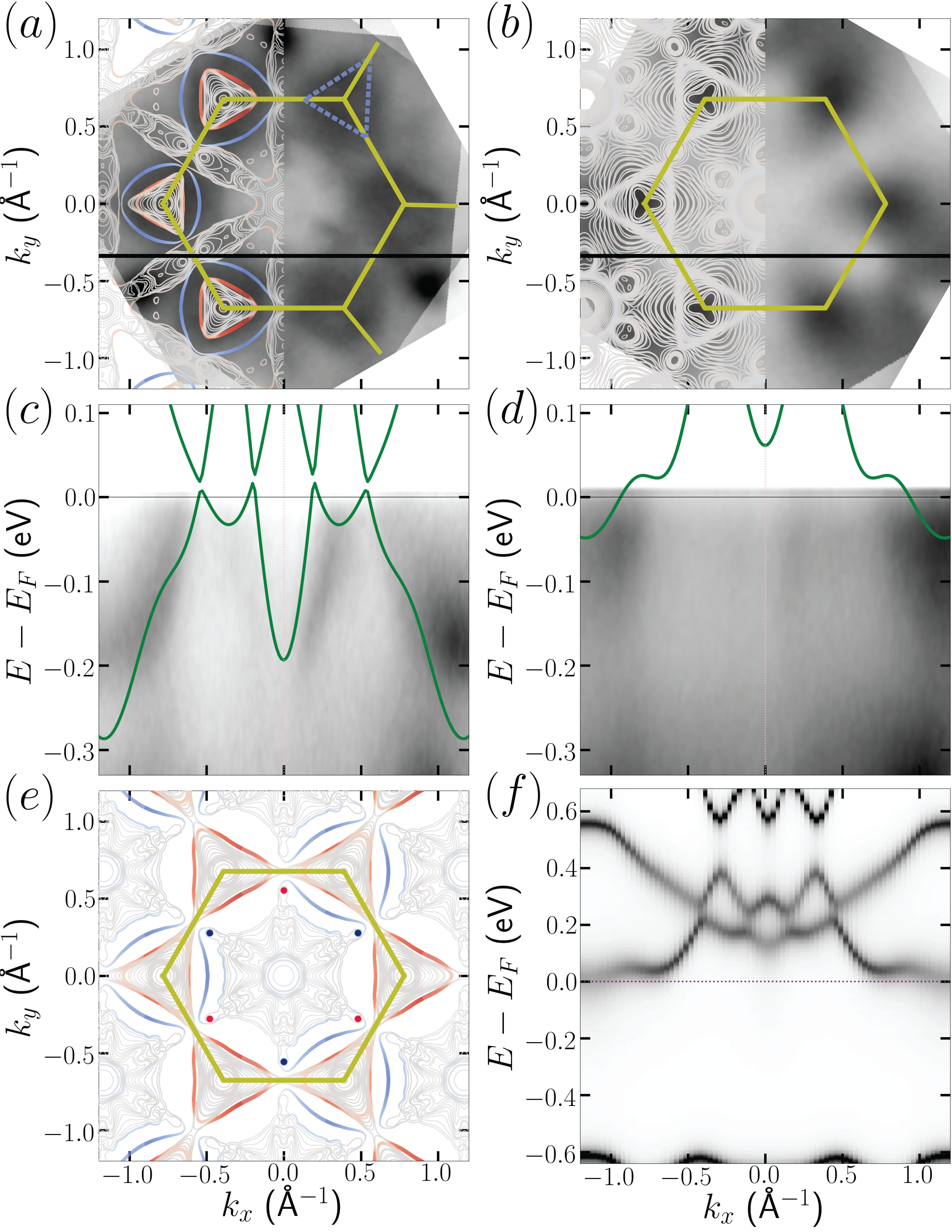}
	\caption{
		ARPES spectra below and above $T_c$ with computed overlays. Constant energy maps of ARPES spectra at $E = E_F$ below $T_c$ ($T=30$K) (a), and above $T_c$ ($T=200$K) (b), overlaid with the Fermi surface contours for the ferromagnetic (a) and non-magnetic (b) slab calculation. Blue dashed lines highlight the location of Fermi arcs in the first and second BZ. For the cuts indicated by the horizontal black lines in (a),(b), the band structure is shown in (c) and (d). The Weyl cones visible in (c) disappear above $T_c$ (d). Band structure calculations for the ferromagnetic (c) and non-magnetic (d) regimes are overlaid for comparison. (e) shows Fermi surface contours computed at the Weyl energy. (f) shows the band structure computed with LDA+DLM.
	}
	\label{fs-overlay}
\end{figure}

Figure \ref{fs-overlay} compares photoemission data for the Sn termination across $T_c$ with the results of numerical calculations. The ARPES constant energy maps shown in Fig. \ref{fs-overlay}(a)-(b) are overlaid with constant energy slices of the bands computed for the slab geometry using the Wannier--basis Hamiltonian extracted from LSDA. These Fermi surface contours are colored based on the probability density contribution at each $\bm{k}$-point; red for the S surface and blue for Sn, as indicated by the corresponding planes in Fig. \ref{structure}. 

For the ferromagnetic calculation (Fig. \ref{fs-overlay}(a)), the computed states have strong contributions from the Sn surface that form petal--like triangular features at the corners of the two-dimensional BZ, which match well with the dark regions of high spectral density seen in the ARPES data. The reason for the apparent six--fold rotational symmetry in both theory and experiment, is that ferromagnetism breaks the $\bm{k} \leftrightarrow -\bm{k}$ symmetry along the $k_z$ direction, which is effectively integrated out for these data, meaning deviations from six--fold symmetry are very slight. Specifically, there is a small anisotropy in the theory calculation between $K$ and $K^\prime$ corners at opposite momenta, consistent with the $C_3$ rotational symmetry.

Above the transition temperature, the Fermi surface looks markedly different (Fig. \ref{fs-overlay}(b)). In the photoemission data, there is a blurring of spectral details stemming from the expected magnetic disorder and fluctuations. Simultaneously, the absence of $\mathcal{T}$--breaking terms in the non--magnetic theory calculation manifests as a six--fold rotational symmetry of the contours, along with a near total absence of surface state contributions at $E_F$.

As the energy is raised above $E_F$ in the ferromagnetic calculation (Fig. \ref{fs-overlay}(e)), the constant--energy slices reveal the topological nature of the surface states. The blue (Sn) surface states pull away from the $M$ points of the BZ with increasing energy, and the three trivial surface states merge with the bulk and vanish. The remaining three Sn surface states connect with the six Weyl cone projections, located near the $M$ points along the $\Gamma-M$ lines, and form topological Fermi arcs as shown in Figure \ref{fs-overlay}(e). To explicitly demonstrate the connection between the Fermi surface and topological Fermi arc states just above $E_F$, in our Supplemental Information we provide an animation of the slab calculation over a range of energies \cite{supplement}.

Having discussed the changes in fermiology across $T_c$, we now examine the details of the bands which form the Weyl cones. We inspect the cut in the $k_x$ direction midway between $K-\Gamma-K^\prime$ and the BZ edge, marked by black lines in Figures \ref{fs-overlay} (a) and (b). This cut passes through the $M$ points of the BZ, and crosses the Weyl cones of the two topological points on either side (Fig \ref{fs-overlay}(c)-(d)). The ferromagnetic LSDA band structure shown in green highlights the two pairs of conical bands that rise upward towards the Fermi energy to form the Weyl cones, similarly to those seen in prior photoemission measurements \cite{liu_magnetic_2019}. While the band connecting the cones in each pair is located above $E_F$ and is not directly visible in the presented ARPES data, the occupied sides of the conical features are clearly visible to the left and right of $k_x = 0$ in the photoemission data (Fig \ref{fs-overlay}(c)). Above $T_c$, the occupied side of the cone--like features are nearly absent from ARPES data, consistent with the LDA+DLM calculation (Fig \ref{fs-overlay}(f)) which shows the correlated state just above $E_F$. The faint spectral features visible near $k_x \sim 0$ and $k_x \sim 1.1$ \AA$^{-1}$ around $-0.3$eV in the photoemission spectra can be explained as the remnants of the lower Hubbard that have not entirely disappeared due to the surviving local magnetic order.  This is consistent with evidence for the persistence of the lower Hubbard band presented in Fig. \ref{fig:TdepGK} (e).

The annihilation of bulk Weyl nodes should be accompanied by changes in the Fermi arcs that ultimately also lead to their disappearance. The Fermi arc features are most clearly seen in the second BZ and are indeed absent above T$_c$.  The $\Gamma$-$K$ cut in Fig. \ref{fig:TdepGK} also cuts through the band that has been identified as the one forming the Fermi arc \cite{liu_magnetic_2019}, and shows a finer temperature-dependence. The band that forms the Fermi arc shifts to momenta closer to $K$ by 135K, and is absent at T$_c$.



\section{VI. Conclusions and Outlook.}

In this work, we have shown how the half-metallic ferromagnetism of Co$_3$Sn$_2$S$_2$ emerges in the partially filled Co-$3d_{x^2-y^2}$ band crossing $E_F$. The critical physics develops on the cluster of Co atoms, for which the highest occupied half--filled state creates a resonance which renormalizes the Coulomb interaction to one-third of its nominal value. The comparable energy scales of bandwidth and interactions result in a strongly correlated magnetic state. As the temperature is raised above $T_c$, the system undergoes a transformation where long range order disappears and the Zeeman--split bands merge into one, transitioning from a Mott ferromagnet to correlated metal. 

The details of the magnetism in this material are highly relevant to its topological properties. However, investigations of this topology in the context of bulk-boundary correspondence need to consider the variety of possible surface terminations in this material and associated differences in surface band structure. The spectroscopic characterization presented here identifies key features that are distinct for the S and Sn terminations, and can be used to differentiate these surfaces. Characteristic signatures of the different surface terminations will be valuable in future experiments, which are needed to understand further the various aspects related to the disappearance of topology, including the role of magnetic domains, local magnetic correlations, and vanishing topological Fermi arcs across the magnetic transition.

The ferromagnetism in Co$_3$Sn$_2$S$_2$ breaks $\mathcal{T}$ symmetry and is a necessary prerequisite for the existence of topological Weyl points. The strongly correlated metallic state in Co$_3$Sn$_2$S$_2$ emerges as a result of magnetic disorder at higher temperatures, restoring $\mathcal{T}$ symmetry and precluding the existence of topology. The detailed behavior of magnetic fluctuations has an effect on topology not just above $T_c$, but throughout a wide temperature range approaching the transition. 
%
In this way the topological properties and strong correlations in this system are intricately linked, and one cannot be adequately considered without the other.  Naively, the Weyl points exist as a result of broken time-reversal symmetry, and should annihilate when the magnetism breaking this symmetry vanishes. However, in the present system, magnetism goes away gradually with temperature as spin-fluctuations increase and the local moments become disordered. This process breaks the local translation symmetry of the crystal, which is necessary for Weyl points to exist. As such, it is not yet clear whether the Weyl points explicitly annihilate or simply cease to exist when the underlying translation symmetry is broken.  This is an important point for future studies to clarify.

~\\

During the review process, two other works \cite{ZHasan2021-Tdependent,YLChen2021-Tdependent} which examine the topological features in Co$_3$Sn$_2$S$_2$ across $T_c$ were uploaded to arXiv.

\section{\textbf{Acknowledgements}}
\begin{acknowledgments}The authors acknowledge helpful discussions with E. H. da Silva Neto and Matthew Staab.  I.V. and A. R. acknowledge support from UC Davis Startup funds and the Alfred P. Sloan Foundation (FG-2019-12170). Theoretical work by V.I. and S.S. was supported by NSF DMR Grant No. 1832728. V.T. and Z. S. acknowledge support from the UC Lab Fees Research Program (LFR-20-653926) and UC Davis Startup funds. This research used resources of the Advanced Light Source, a U.S. DOE Office of Science User Facility, under contract no. DE-AC02-05CH11231. 
\end{acknowledgments} 

\bibliography{references}

\end{document}